\documentclass[%
 aip,
 amsmath,amssymb,
 reprint,%
]{revtex4-1}
\usepackage{graphicx}% Include figure files
\usepackage{dcolumn}% Align table columns on decimal point
\usepackage{bm}% bold math
\usepackage{xcolor}
\usepackage[utf8]{inputenc}
\usepackage[T1]{fontenc}

\usepackage{mathptmx}
\usepackage{gensymb} % needed for different symbols in text (like degrees)
\begin{document}
\preprint{AIP/123-QED}
\title{Force spectroscopy with electromagnetic tweezers}
% Force line breaks with \\
\author{Joseph G. Piccolo}
\affiliation{
Department of Physics, Emory University, 400 Dowman Dr., Atlanta, GA 30322}
\author{Joshua M\'endez Harper}%
\affiliation{Department of Earth Science, University of Oregon, 1272 University of Oregon, Eugene, OR 97403}
\author{Dan Kovari}
\author{David Dunlap}
\affiliation{
Department of Physics, Emory University, 400 Dowman Dr., Atlanta, GA 30322}
\author{Laura Finzi }
\homepage{http://www.physics.emory.edu/faculty/finzi/}
\affiliation{
Department of Physics, Emory University, 400 Dowman Dr., Atlanta, GA 30322}
\date{\today}% It is always \today, today,
\begin{abstract}
Force spectroscopy using magnetic tweezers (MT) is a powerful method to probe the physical characteristics of single polymers. Typically, molecules are functionalized for specific attachment to a glass surface at one end and a micron-scale paramagnetic beads at the other. By applying an external magnetic field, multiple molecules can be stretched and twisted simultaneously without exposure to potentially damaging radiation. The majority of MT utilize moving permanent magnets to produce the forces on the beads (and the molecule under test). However, translating and rotating the permanent magnets may require expensive precision actuators, limits the rate at which force is changed, and may induce vibrations that disturb tether dynamics. Alternatively, the magnetic field can be produced through an electromagnet which allows much faster force modulation and eliminates motor-associated vibration. Here, we describe a low-cost quadrapolar electromagnetic tweezer design capable of manipulating DNA-tethered MyOne paramagnetic beads with forces of up to 20 pN. The solid-state nature of the generated B-field modulated along two axes is convenient for accessing the range of forces and torques relevant for studying the activity of DNA motor enzymes like polymerases and helicases. Our design specifically leverages technology available at an increasing number university maker spaces and student-run machine shops. Thus, our design is not only applicable to a wide range biophysical research questions, but also an accessible tool for undergraduate education.
\end{abstract}
\maketitle
\section{\label{Intro}Introduction}
\subsection{\label{MagTweBio}Magnetic tweezers in biophysical research}
Magnetic tweezers (MT) comprise a class of force spectroscopy tools that are well-suited to the characterization and manipulation of individual biomolecules, polymers, and even live cells \cite{sarkar2016guide}. Electromagnets were first presented by Strick et al. \cite{strick1996elasticity} to study the elasticity double-stranded DNA. Since then, investigators have used MT to experiment on modified and unmodified nucleic acid polymers in buffers with and without small molecules and salts that modify conformations, as well as to examine the interaction of these polymers with processive enzymes and other proteins. 
In typical magnetic tweezer studies, single-molecules of DNA are functionalized for specific attachment to a glass surface at one end and a micron-scale paramagnetic bead at the other. The Brownian motion of the tethered bead is confined by the DNA tether anchoring it to a point on the glass. Monitoring the position of the bead as a function of time reveals the anchor point and the effective length of the tether connecting the bead. The attached paramagnetic bead experiences a force proportional to the gradient of the applied external magnetic field.\cite{MThandbook2009} Modulation of a sufficiently strong field exerts attractive force and torque on the paramagnetic bead (Fig. 1A). 
The utility of magnetic tweezing lies in its ability to arbitrarily manipulate DNA topology in real-time. This is especially relevant because DNA is normally topologically constrained \textit{in vivo} via tension and twist, conditions that are exceedingly difficult for experimentalists to recreate biochemically.\cite{Travers2015, Roca2011}  When mild attractive force draws the bead away from the anchor point, the DNA tether stretches like a spring exerting entropic force. Torque applied to rotate the bead twists  the molecule, imparting supercoiling. When supercoiling exceeds a critical threshold, the molecule buckles and writhes into a plectonemic form, shortening overall tether length. Thus, changes in the effective tether length are the readout for topological manipulations of DNA via tension and twist (Fig. 1B). 
The breakthrough associated with MT in biophysical research stems from the fact that DNA processing enzymes like polymerases impart tension and twist within particular force and torque regimes. Using this tool, researchers have investigated not only DNA mechanics but also the activity of DNA topoisomerases\cite{Charvin2003}, helicases\cite{Seol2016}, and RNA polymerases\cite{Revyakin2004}, contributing to our understanding of their respective mechanisms.
\subsection{\label{FerroTwee}Permanent, rare-earth magnetic tweezers}
Typical magnetic tweezers have a pair of permanent, rare-earth magnets near the sample. Field strength at the sample plane is modulated by mechanically moving the magnets closer to or further from the sample. Similarly, field orientation is controlled by rotating the magnets. Although this design generates the necessary magnetic forces, it has a few drawbacks. The motors, which physically move the permanent magnets, may induce vibrations. Such disturbances limit the resolution with which tether length and force can be measured. Additionally, the sample is usually illuminated by a beam passing through a narrow gap between the magnets. Thus, the magnets must be carefully aligned with the objective to avoid variations in illumination strength as the magnets rotate or translate. Lastly, the magnetic field may be axially asymmetric, causing a tethered bead to precess as the magnets are turned. To ameliorate these effects, most permanent magnetic tweezers are operated using only integral turns of the magnetic field to twist the tether.
\begin{figure}
\includegraphics[scale=0.5]{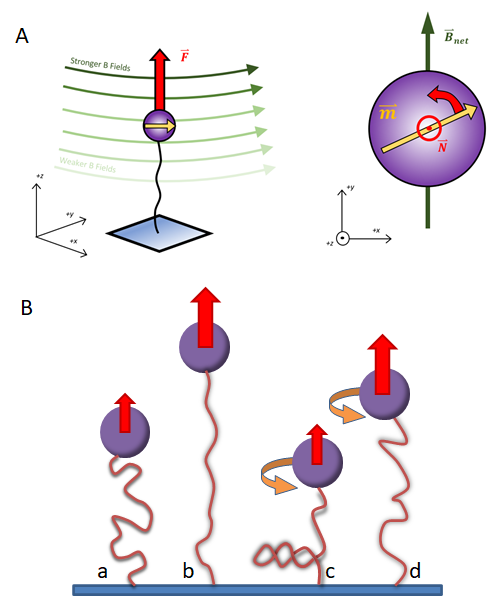} % Here is how to import EPS art
\caption{\label{fig_1} \textbf{Magnetic Bead Field Interactions:} \textbf{(A)} A molecule of DNA is shown immobilized to a glass surface (blue) and attached to a magnetic bead (purple). The magnetic moment of the bead (yellow) aligns itself with the external magnetic field (B, green) propagating from permanent magnets above (not shown) in the +z direction. \textit{Left} The gradient of the B field attracts the bead along \textit{z} (red). \textit{Right} A top-down view. Torque is generated when the magnetic field (green) rotates about a normal to the xy-plane. If there is no swivel in the attachment of the DNA to the bead, the DNA tether is twisted causing supercoiling. \textbf{(B)} The conformation of DNA depends on tension and torque. \textit{(a)} An untwisted molecule under low tension \textit{(1)}  extends as the tension increases \textit{(b)}. Alternatively, twisting the DNA under low tension supercoils the molecule into plectonemic loops, reducing the tether length \textit{(c)}. Raising the tension on the highly supercoiled molecule converts the writhe back to twist as the DNA loops are stretched. \textit{(d)}.} 
\end{figure}
\section{\label{EmagTwee}Electromagnetic tweezers}
The limitations associated with permanent magnets may be overcome by replacing the mechanically-driven magnetic system with an electromagnetic one. Electromagnetic tweezers use a set of solenoids to generate the magnetic field. The field strength and orientation is modulated by changing the current through the solenoids. Such modulation can be more rapid than that achieved by moving permanent magnets to alter that magnetic field strength and orientation. Previous implementations of magnetic tweezers have been reported. A hexagonal pattern of solenoids with Mu-metal cores was described which produced forces on super-paramagnetic beads as high as 20 pN vertically and 5 pN horizontally and rotations at 10 Hz. The large potential well of this system even allowed horizontal positioning of 5 micron-diameter magnetic beads\cite{Gosse2002}. In another implementation, the recording head of an audio tape device was employed to assemble a rotationally static tweezer that produced up to 50 pN with 1 A of current and allowed high bandwidth (10 kHz) force modulation\cite{Tapia-Rojo2018}. A tweezer with four laterally set solenoids generating a horizontal field for rotation and another axially placed solenoid adding a vertical component generated fields as high as 8 pN with 3 A of current.\cite{Jiang2016}. A similar instrument with separate rotational and axial control was formed of Helmholtz coils that create rotational stiffness of several pN$\cdot \mu$m/rad and an axially placed permanent magnet to apply tension up to tens of pN  \cite{Janssen2012}. A combination of eight lateral solenoids with an axial optical trap can produce much larger forces and torques and boasts very high bandwidth force measurement \cite{Sacconi2001}. Clearly, eliminating the risk of radiative damage to biomolecules and the multiplexing possibilities make purely magnetic tweezers attractive.

Here, we describe a simplified electromagnetic tweezer design using two pairs of solenoids in an orthogonal arrangement. We demonstrate that this implementation can manipulate MyOne parametric beads with forces of over 10 pN. This instrument leverages manufacturing equipment that is increasingly available in university makerspaces and student machine shops. Thus, our eMT not only provides a system to manipulate a wide range of biopolymers and associated enzymes, but also a robust teaching tool in biophysical education.

\section{\label{reqs}Design requirements}
\begin{figure}
\includegraphics[scale=0.4]{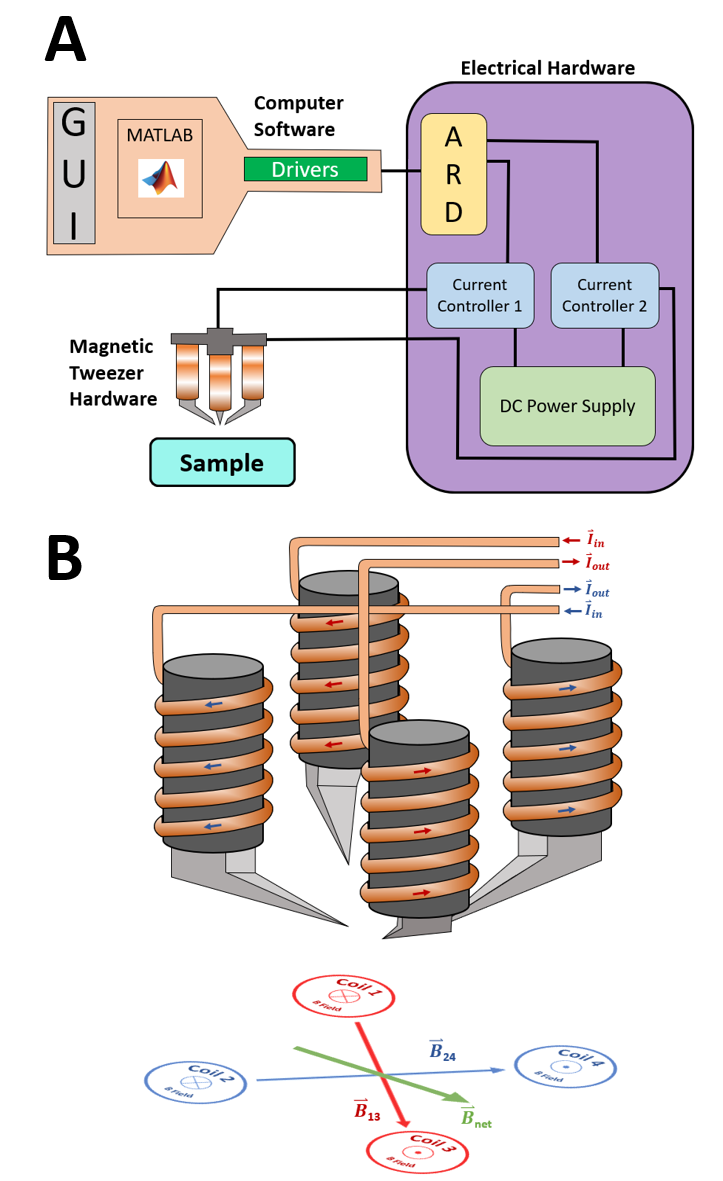}
\caption{\label{fig2} \textbf{System Summary} \textbf{(A)} Three main components were integrated for the electromagnetic tweezer. A GUI developed in MATLAB issued commands through drivers interfaced through a serial port to custom electrical circuitry that via two controllers distributes current from a DC power supply to solenoids at opposite corners of the electromagnet frame. \textbf{(B)} Coils on opposite corners of the square pattern are wired in series to align the fields between the pole pieces that extend below. Modulating the magnitude and polarity of currents through these pairs of solenoids generates fields (red or blue arrows) that combine vectorially to create a net magnetic field (green). } 
\end{figure}

The essential components for a set of electromagnetic tweezers are an electromagnet that produces a highly non-uniform magnetic field near a sample of interest, controlling electronics to modulate the strength of the field, optical components capable of resolving tethered paramagnetic beads, and software to track bead motion. A block diagram of the system is rendered in \textbf{Figure \ref{fig2}}.

As an instrument for undergraduate education in biophysics, we attempted to devise an eMT that met the following criteria. Firstly, the construction of the apparatus utilized components and/or manufacturing tools generally available within physics and engineering departments. For instance, the design of the electromagnet soft-iron cores were constrained by whether the geometries could be machined in their entirety using manual lathes and mills commonly found in student machine shops or maker spaces. Secondly, the electrical control hardware was based on popular open-source physical computing platforms that involve low barriers to entry, such as Arduino. In addition, we ensured that any custom electrical hardware for this project could be manufactured using a number of different techniques based on the resource availability. Lastly, as budgets for undergraduate projects are often limited, we sought to build an instrument with relatively low cost. Overall, we estimate that this instrument can be constructed for under 500 US dollars. 

An equally important constraint involves the forces needed to interrogate biomolecules and their interactions with proteins. The dynamics of a single paramagnetic bead (PMB) with mass $m_b$ and velocity $\textbf{u}_m$ in a liquid exposed to an magnetic field gradient is given by:

\begin{equation} \label{motion}
m_b \frac{d\textbf{u}_b}{dt} = \textbf{F}_d + \textbf{F}_M,
\end{equation}

where $\textbf{F}_D$ is the Stoke's drag and $\textbf{F}_M$ is the magnetic force. The drag force $\textbf{F}_D$ on a bead with radius $r$ suspended in a mediums with dynamic viscosity $\eta$ is:

\begin{equation}
\textbf{F}_D = - 6 \pi \eta r \textbf{u}_b
\end{equation}

The magnetic force on the PMB depends on the negative gradient of the magnetic energy $U$:

\begin{equation} \label{force}
\textbf{F}_M = - \nabla U  = \frac{1}{2} \nabla (\textbf{m}(\textbf{B}) \cdot \textbf{B})
\end{equation}

In equation \ref{force}, $\textbf{B}$ is the magnetic flux density, $\textbf{m}(\textbf{B})$ is the magnetic moment of the PMB (which generally  depends on the magnetic field). For small relatively weak fields, the force on the PMB scales linearly with gradient of the square of the magnetic flux density and the difference in magnetic susceptibility $\Delta \chi = \chi_b - \chi_a$ between the bead material and ambient medium (one can often neglect the ambient susceptibility because it tends to be several orders of magnitude smaller than that of the PMB):

\begin{equation} \label{force}
\textbf{F}_M = \frac{V_b \chi_b}{2 \mu_0} \nabla |\textbf{B}|^2,
\end{equation}

where $V_b$ is the bead's volume and $\mu_0$ is the permeability of free space. For larger fields, conversely, the magnetic moment of the beads reaches a  saturation value $\textbf{m}_\text{sat}$ and the expression for magnetic force reduces to:

\begin{equation} \label{magForce}
    \textbf{F}_M  = \frac{1}{2} \nabla (
 \textbf{m}_\text{sat} \cdot \textbf{B}).
\end{equation}

Micron-sized beads reach terminal velocity very rapidly in the viscous media and the acceleration term on the left-hand side of equation \ref{motion} goes to zero. Thus, the motion of a bead is governed by the balance between the magnetophoretic force and viscous drag:

\begin{equation} \label{}
6 \pi \nu r \textbf{u}_b = \frac{1}{2} \nabla (
 \textbf{m}_\text{sat} \cdot \textbf{B}).
\end{equation}

\begin{figure*}[t]
\includegraphics[scale=0.3]{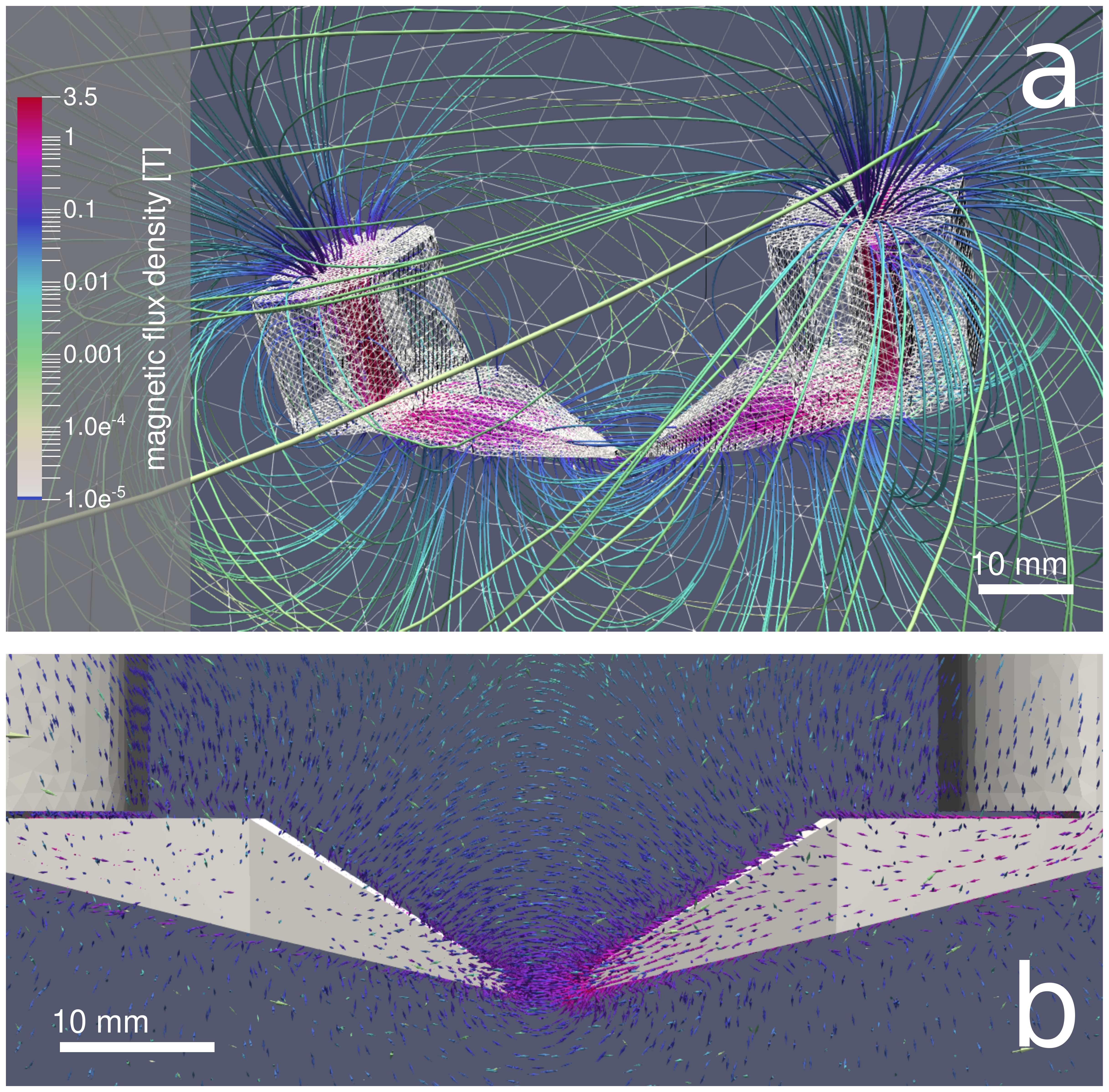}
\caption{\label{model} \textbf{Finite element modeling.} We simulated the performance of the electromagnetic tweezer using ElmerFEM, an open-source finite element analysis package. Rendering was accomplished in Paraview. \textbf{(a)} Magnetic flux density streamlines under operation of a single pole pair with 1 A flowing through each solenoid. \textbf{(b)} Detail of magnetic flux density in the vicinity of the pole tips under the same conditions as above.}
\end{figure*}

\begin{figure}
\includegraphics[scale=0.6]{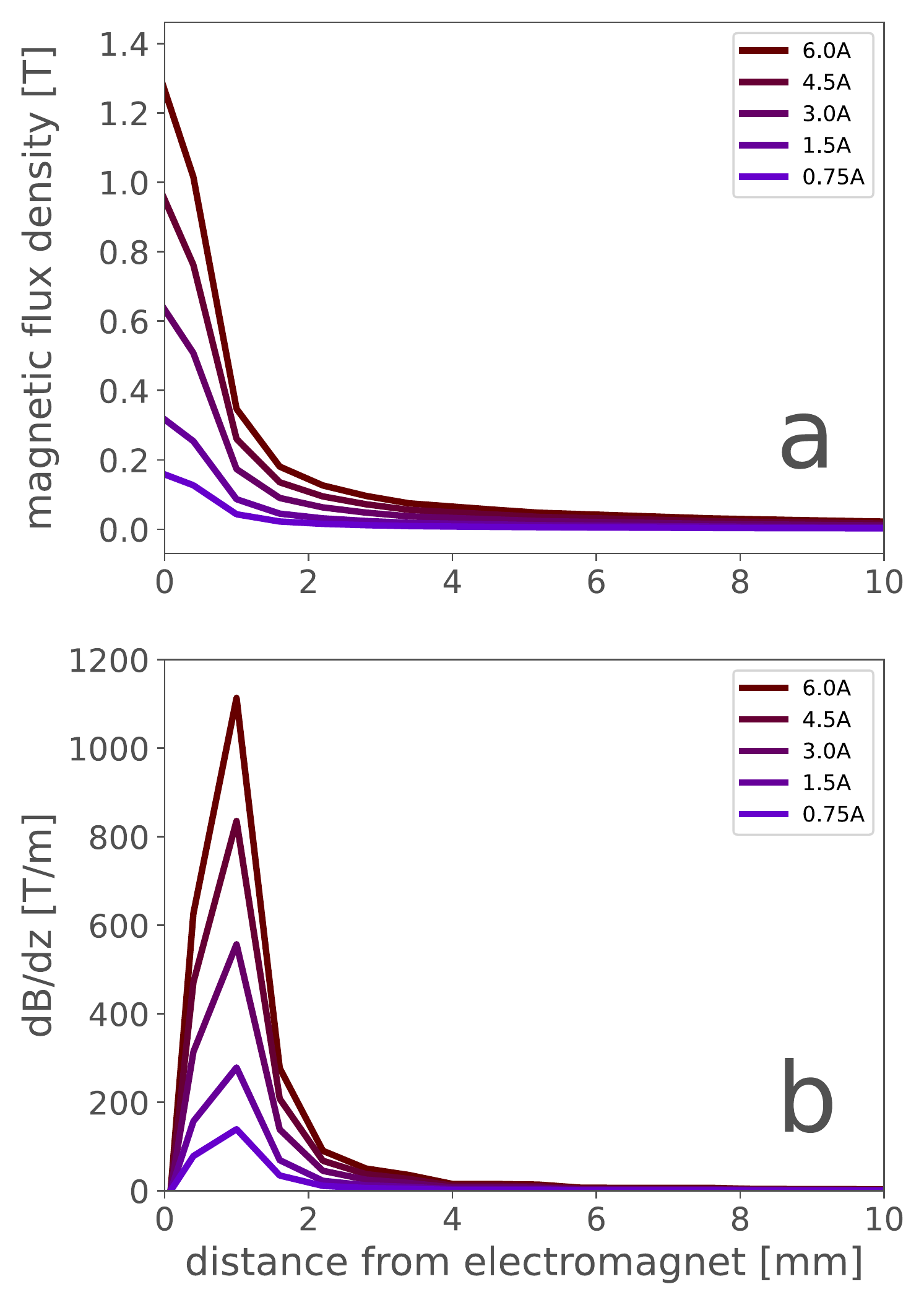}
\caption{\label{mag} \textbf{Magnetic field parameters} \textbf{(a)} Magnetic flux density as a function of distance from the pole pieces shown in Figure \ref{model}.  \textbf{(b)} Vertical gradient of the magnetic flux density as function of distance from pole pieces. Force peaks near 1 mm. } 
\end{figure}

Implicit in equation \ref{magForce} lies the fact that the force on the bead depends on the magnetic field gradient, not the magnitude. Thus, an effective eMT requires an electromagnet capable of producing a highly non-uniform magnetic field. For instance to obtain trapping forces on the order of tens of pico-Newtons on 1 $\mu$m Invitrogen MyOne requires magnetic flux density gradients in the range of 100s to 1000s Tm\textsuperscript{-1}.

The torque $\tau$ is given by:

\begin{equation} \label{}
    \tau =\overrightarrow{m_o} \times \overrightarrow{B}.
\end{equation}

Above, the $m_o$ is a component of magnetic moment not aligned with the external field. Unlike the force, $\tau$ depends on the magnitude of the magnetic flux density.

We investigated the performance of various configurations of the four pole pieces using finite element analysis (FEA). Specifically, we simulated our system using Elmer, an open-source solver developed  CSC - IT Center for Science (CSC). Meshing for the model was done using gmsh (another open-source tool) which allows for flexible mesh refinements, especially in regions where we expected high field gradients. Simulations where conducted in full 3D and results were visualized in Paraview (Figure 3).

The magnetic tweezer includes four independent solenoid coils (460 loops each) wrapped around steel cores. Magnetic fields are compressed along tapered, custom-machined pole-pieces (see Figure 3), so that large gradients and high forces can be achieved with relatively low currents (Figure 4). Using the open-source finite-element modeling software Elmer, we estimate that the strength of the field from the electromagnet with 1.4 Amps of current should equal that of a permanent magnet used in previous work. By  controlling  each  solenoid  current independently, the magnetic field can be aligned/positioned arbitrarily, without the vibrations or changes in illumination that affect particle tracking. 

Four solenoids arranged vertically on the four corners of a square frame, with coils on opposing corners wired in series (Fig. 2B). Each current controller diverts specified levels of current from an external DC power supply to one of the pairs of solenoids. By varying the electrical current through each pair of solenoids, the vector sum of the resulting \textit{B} field can be modulated and rotated 360\degree in the \textit{xy} plane.

\begin{figure}
\includegraphics[scale=0.6]{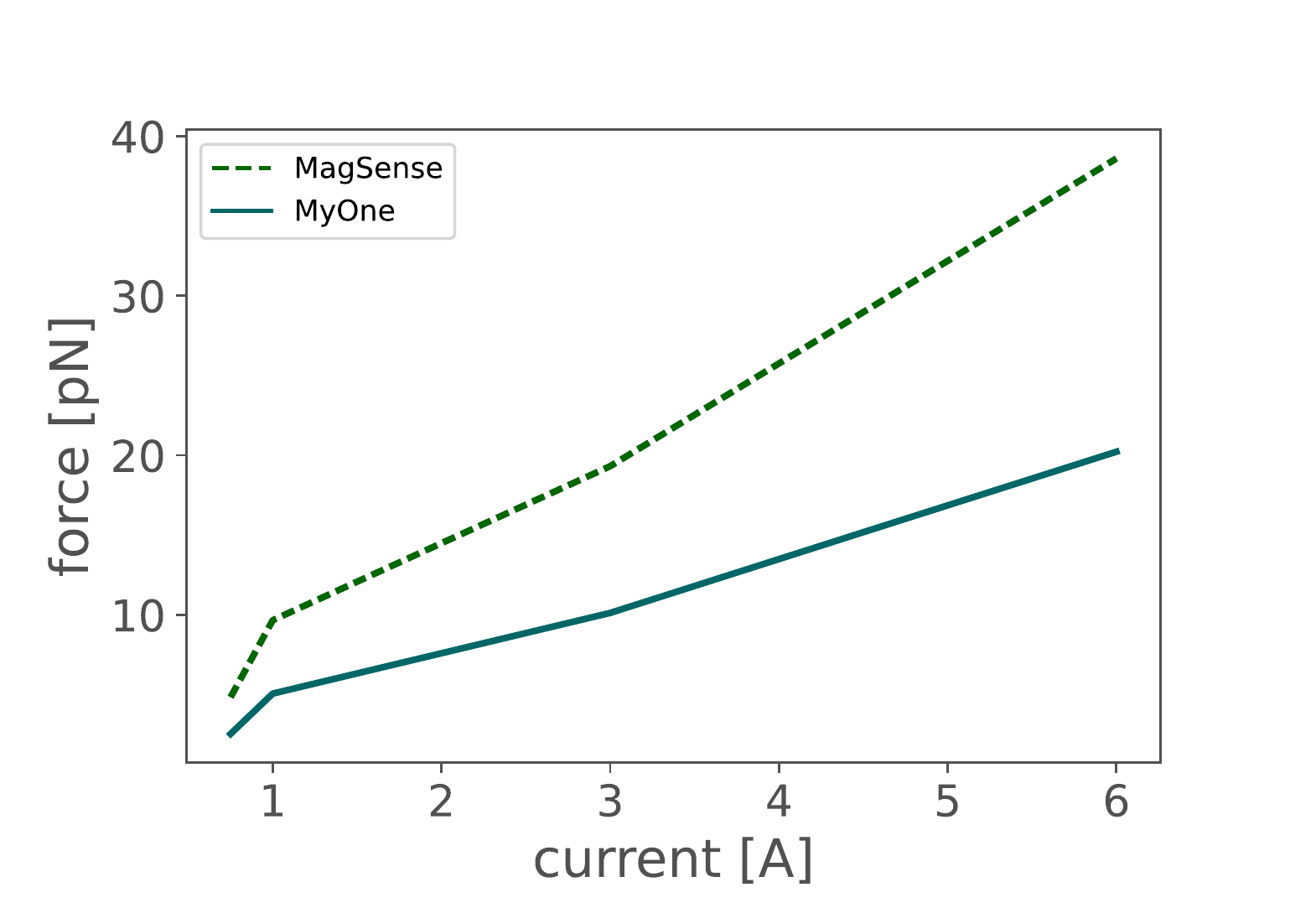}
\caption{\label{mag} \textbf{Magnetic field parameters} \textbf{(A)} Magnetic flux density as a function of distance from the pole pieces shown in Figure \ref{model}.  \textbf{(B)} Vertical gradient of the magnetic flux density as function of distance from pole pieces. Force peaks near 1 mm. } 
\end{figure}

\subsection{Electrical hardware}

\subsubsection{eMT driver board}

Unlike conventional, permanent magnet electromagnetic tweezers, eMTs do not require moving parts or mechanical actuators to modulate the magnetic field in the vicinity of a supraparamagnetic bead. The field is adjusted by changing the magnitude and/or direction of current through the pole-piece solenoids. For the configuration simulated above, forces on the order of 10s of pico-Newtons require maximum solenoid currents between 3-6 A (Figure 5).  Controlling currents of this magnitude can be achieved easily using conventional H-bridge circuits. Because H-bridges are employed widely to drive inductive loads (primarily DC motors), they are readily available as monolithic integrated circuits (IC) capable of driving large currents. Our eMT uses two ST VNH5019 automotive H-bridges, which can deliver continuous currents of 12 A (with appropriate heat sinking) and 30 A peak current. The VNH5019 has been adopted by the hacking and maker communities because it can easily interface with a popular microcontrollers through 2.5-5 V logic levels and pulse-width modulation.

\begin{figure*}[t]
\includegraphics[scale=0.15]{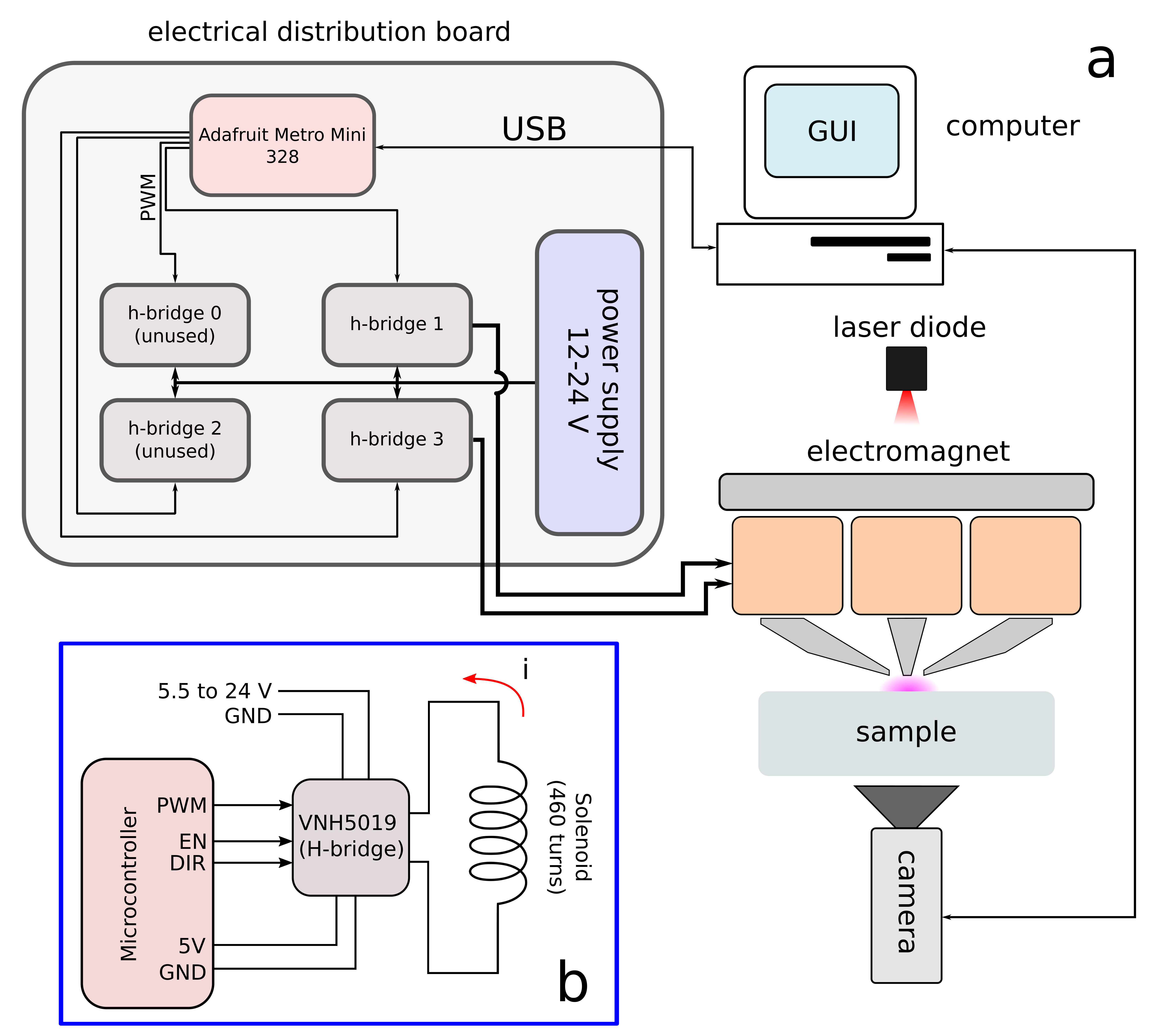}
\caption{\label{schem} \textbf{Electrical block diagram.} Schematic representation of a single solenoid drive circuit.} 
\end{figure*}

A schematic representation of the eMT driver circuit is shown in \textbf{Figure \ref{schem}}. To provide maximum flexibility, our design implements 4 H-bridge ICs which can be configured in two manners. In the first configuration, each VNH5019 drives a single solenoid coil independently. This allows the user to compensate for any operational mismatch between solenoids. The second configuration connects opposing pole pieces in series to a single H-bridge IC. In this last mode, the two unused H-bridge IC's may be used as power cooling devices to prevent overheating of the solenoid hardware and sample under test.

We designed the printed circuit board (PCB) using KiCAD (files available in Supplementary Material), an open-source electronic design automation suite available for Linux, Mac, and Windows (https://www.kicad.org/). KiCAD supports multilayer- board (PCB) designs, meaning that electrical traces can be routed on both the top and bottom sides of the board as well as internally through the board. For the present work, we designed a four-layer board where the top and bottom layers served to route power and signals. The two internal layers consisted of large copper pours that funneled heat from the H-bridges to the circuit board's enclosure via aluminum standoffs. This printed circuit board was produced through a professional fabrication house offering student discounts (Advanced Circuits).

While some investigators may not have access to professional PCB fabrication services, PCBs may be produced ``in-house" through a variety of techniques (albeit most of these procedures only allow 1 or 2-layer designs). We determined that our design could also be implemented using a PCB milling machine (2 layers) and chemical etching (1 layer). Since both of these procedures do away withexclude the internal thermal layers they may limit the amount of current.

A custom-built reflow oven and through-hole soldering were used to complete the circuitry, which was enclosed in a clear plastic casing to prevent shorts and facilitate cabling. 

\subsection{Software}

An intuitive GUI was built to command the microcontroller via USB and control the magnitude and polarity of the current applied to the solenoids. Actual control of the coil current is established through an intermediary driver board which receives a pulse-width modulation (PWM) signal from the microcontroller that it translates into the effective power delivered to a pair of coils. An additional digital flag (5V or 0V) specifies the direction of current flow through the pair. The electromagnet control may be operated independently or integrated into other software systems via Matlab's object-oriented structure.

\subsubsection{\label{comp}Components of the prototype}

Additionally, a custom graphical user interface (GUI) for control was developed as a MATLAB script. Through this interface, the user sets the magnetic field strength as well as the \textit{xy} orientation of the field. The MATLAB script in turn passes commands through a device driver to a serial port connected to custom electrical circuitry. The circuitry featured a commercially available microcontroller (Arduino Micro development board hosting an Atmega328 Atmel microprocessor) to regulate current controllers for the orthogonal pairs of the 
 
\section{Experimental}
\begin{figure*}[t]
\includegraphics[scale=0.4]{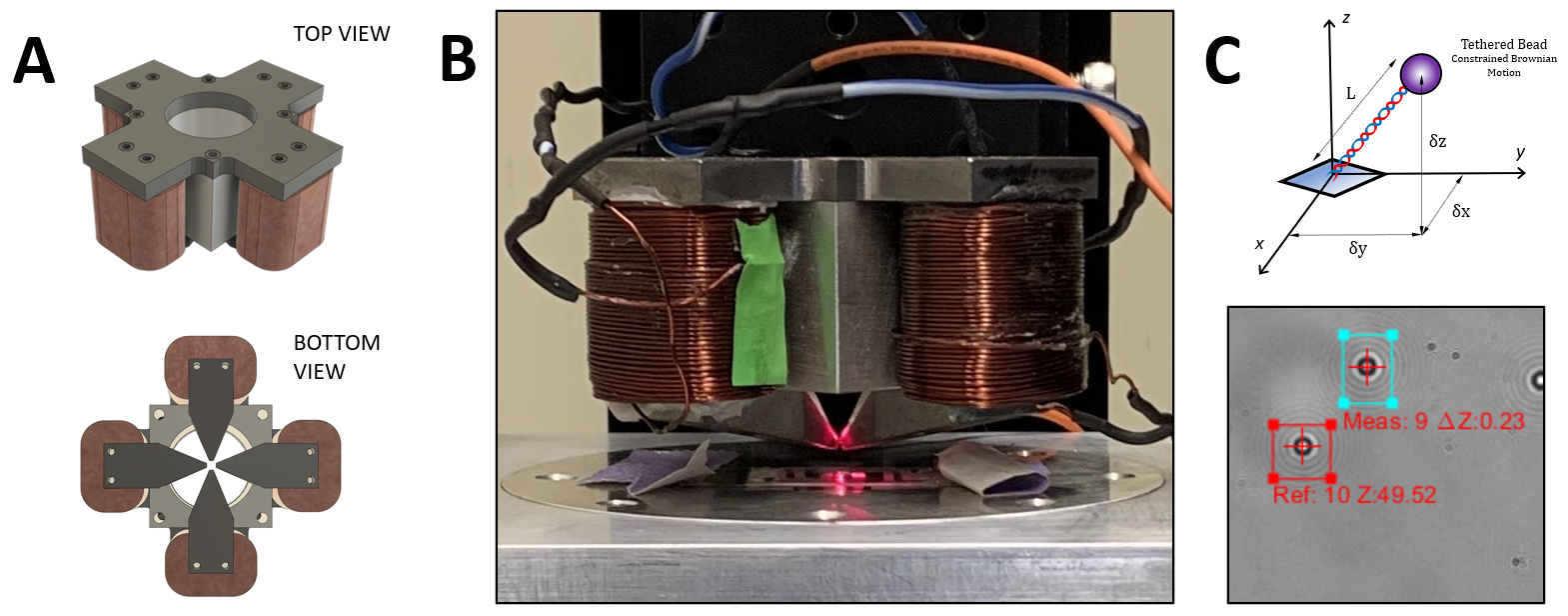}% Here is how to import EPS art
\caption{\label{fig3} \textbf{Electromagnetic Tweezer prototype} \textbf{(A)} This oblique top view illustration (Fusion 360 CAD) of the electromagnetic tweezer shows a low carbon steel frame (dark grey) on which is mounted an aluminum block with a cylindrical central bore (light gray) and four wire solenoids surrounding low carbon steel cores. The bottom view shows octahedral pole-pieces, made from the same low carbon steel, mounted on each solenoid. The tapered ends increase the field gradients and converge to superimpose at the center. \textbf{(B)} In a side view of the prototype in position above the specimen stage of the microscope, 635 nm light from an LED (not pictured) along the optical axis  reflects from the pole pieces and a microchamber taped to the stage. A 63X objective below the sample broadcasts an image through a relay lens onto the camera. \textbf{(C)} \textit{Top} In this diagram of the DNA-tethered bead, the DNA is anchored to the glass at the origin of a Cartesian coordinate system and the coordinates \textit{x}, \textit{y}, and \textit{z} represent the position of the bead. \textit{Bottom} A representative image shows the circular diffraction patterns of DNA-tethered MyOne beads. The \textit{x} and \textit{y} coordinates are established by locating the convergence of local intensity gradient vectors \cite{Parthasarathy2012}, while \textit{z} is calculated from a correlation of diffraction patterns versus displacements of the objective \cite{kovari2019model}.}
\end{figure*}
\subsubsection{\label{partTrack} Particle tracking, tether length, and tension }
Most magnetic tweezers are based on optical microscopy with real-time or offline analysis of diffraction patterns to track beads and determine tether lengths. Light and beads of similar scale, generate circular diffraction patterns of the beads in images (Fig. 3C) that varies according to the distance of the bead from the focal plane. Our magnetic tweezers utilize image analysis to locate points of radial symmetry and determine \textit{x} and \textit{y} image coordinates for each bead. Time averages of these coordinates identify the anchor point. Then the \textit{z} position is determined by determining the best match between the radial intensity profile of the diffraction pattern and those recorded previously for different focal offsets of the objective.\cite{vanLoenhout2012,MThandbook2009, kovari2019model} 
For a time interval with \textit{n} captured images, the average effective tether length \( \langle L \rangle \) is a function of the recorded \(\delta\)x, \(\delta\)y, and \(\delta\)z bead positions of each frame,
\begin{eqnarray}
\langle L \rangle = \frac{1}{n}\sum_{i=1}^{n} (\delta x_i^2 + \delta y_i^2 + \delta z_i^2)^{-1/2}.
\end{eqnarray}
Based on the equipartition theorem, the force on the bead can be calculated as a function of the effective tether length \cite{MThandbook2009, sarkar2016guide} according to
\begin{eqnarray}
\langle F \rangle = \frac{k_BT\langle L \rangle }{\langle \delta y^2 \rangle}.
\end{eqnarray}
\subsubsection{\label{sampPrep} Sample Preparation}
To test the electromagnet, a microchamber with DNA-tethered beads was prepared. To assemble a microchamber, a laser-cut, parafilm gasket with a central opening slightly longer than 22 mm was placed between cleaned 24x50 and 22x22 (mm) coverglasses, and the assembly was heated briefly to seal it. The approximately 15 $\mu$l chamber created by the cavity in the gasket between the glasses was rinsed with phosphate-buffered saline (PBS) and incubated with 4 $\mu$g/mL anti-digoxigenin (Roche, Madison, WI) in PBS for 1 hour. After incubation the chamber was rinsed with PBS.
The DNA construct was composed of a 3837 bp central fragment between a 695 bp biotin- and a 749 bp digoxigenin-labeled tail fragment at opposite ends. The central fragment was amplified using the polymerase chain reaction (PCR) with a plasmid (pDD1N2BbvCI) template and primers containing BsaI or BspQI restriction sites. Biotin- and digoxigenin-labeled fragments were produced in similar PCR reactions with the same plasmid in which 10 percent of the dATP nucleotide was replaced by biotin- or digoxigenin-labeled dUTP. The three fragments were digested with BsaI and/or BspQI (New England Biolabs) and the purified products were ligated with T7 ligase.
In parallel, an aliquot of streptavidin-coated, 1 micron-diameter magnetic beads (MyOne Streptavidin T1, Invitrogen, Grand Island, NY, USA) was washed 2X with 1 molar NaCl and mixed with the DNA construct in 100 mM NaCl binding buffer. Following a 5 minute incubation, the DNA-bead mixture was added to a pre-assembled flow-chamber and incubated for 10 minutes. Unbound DNA and beads were flushed out with 100 mM NaCl stretching  buffer before brief storage at 4\degree C or observation on the microscope.
\section{Results and Discussion}
For each test, the electromagnet pole pieces were positioned at a specified height above the top coverslip and the power supply was set to 12 V. A field of view was selected that included both tethered beads exhibiting constrained Brownian motion and beads stuck to the glass. After selecting small regions of interest for individual particle tracking, a set of ROI images was recorded under high force, aligned, and averaged for each bead at a series of positions of the objective, which was mounted on a piezoelectric positioner. During the experiments with the objective position held constant, this calibration scan served as a look-up-table to determine \textit{z} based on changes in the diffraction pattern. To characterize the force and torque regimes of the electromagnet, different tests were performed.
\subsubsection{\label{StretchDNA} Stretching the DNA tether}
A force extension experiment was used to determine the range of forces generated by the electromagnet. At at height of 1.5 mm with a constant orientation of the magnetic field, the magnitude was varied between 0 and 100\% in steps of 10\%. Beads were tracked for 10 seconds at each step. As expected for a worm-like chain, Fig. 8 (lower) shows that the compact, randomly coiled DNA extended by large increments in response to current changes at low levels. Beyond that, with the DNA tether more extended, similar steps of the current produced smaller length increases.  Fig. 8 (upper) shows that as the current increased from 0 to 100\%, the force exerted on the bead rose from slightly below 1 up to 18 pN  (blue). Little hysteresis was observed as the power was reduced again to zero (red). The force was calculated using the equipartition theorem expression described by Vilfan et al.\cite{MThandbook2009} 
\begin{figure}[!ht]
\includegraphics[scale=0.6]{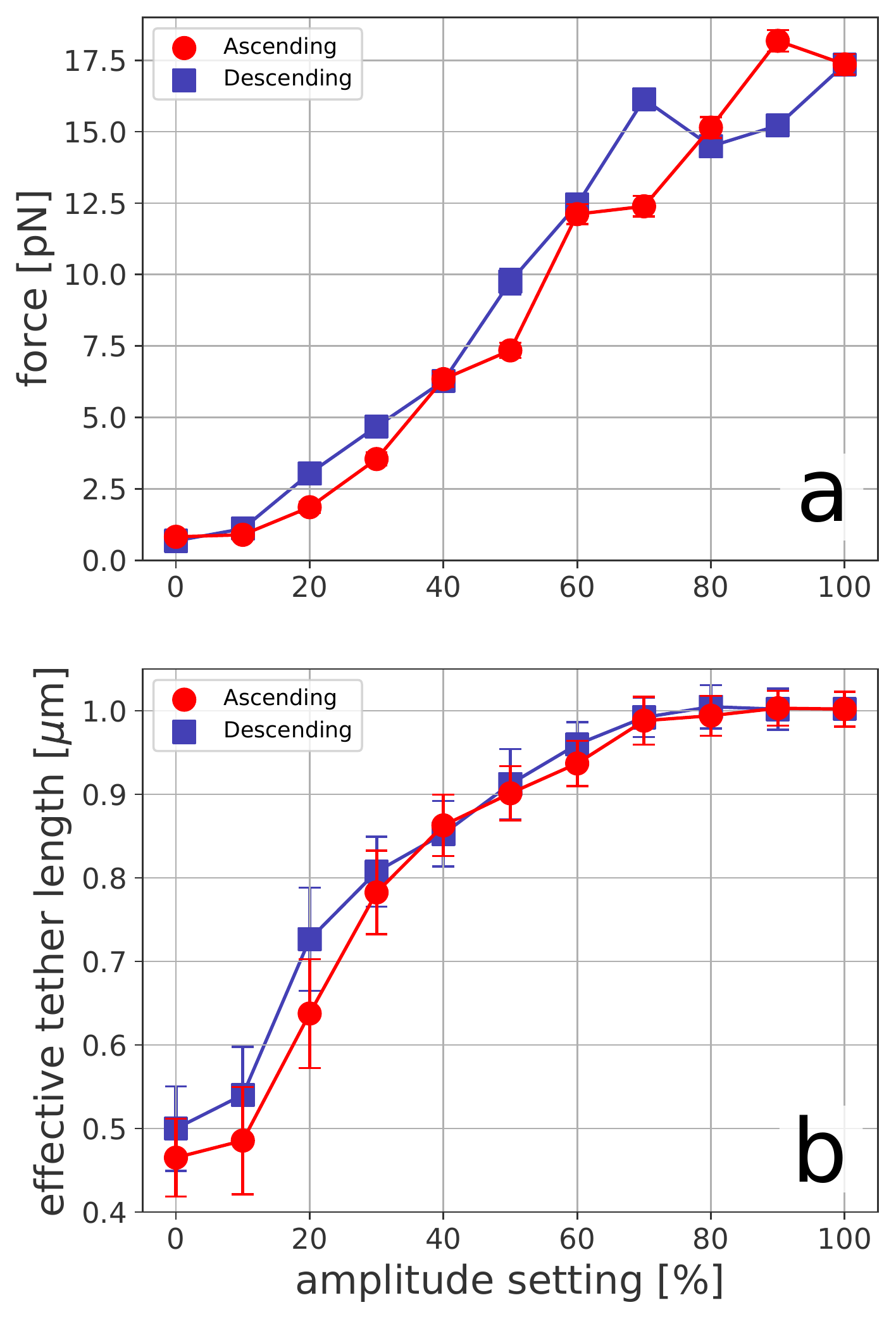} % Here is how to import EPS art
\caption{\label{fig_4} \textbf{Force Extension Results:} \textit{Top} Attractive force (piconewtons) versus current displays a fairly linear relationship with a maximum force of approximately 18 pN.  The error bars are calculated as a function of standard deviation of tether length and \textit{y} variance. \textit{Bottom} The DNA stretched in response to increasing force generated by the electromagnet with negligible hysteresis between DNA stretching (blue) and relaxation (red) sequences. The error bars are the standard deviations in tether length during each time interval. The \textit{x} axis for both graphs is the strength setting at 0\degree orientation with 100\% corresponding to the maximum PWM value 255.} 
\end{figure}
\subsubsection{\label{TwistDNA} Twisting the DNA tether.}
Under low tension, rotating a tethered bead twists the attached DNA and increases the torque in the DNA tether. Additional twist causes the molecule to curl and form a loop and finally a plectoneme, which reduces the overall extension of the DNA. For this experiment, the pole pieces were positioned approximately 2.5 mm above the top coverslip, and the current (force) of the electromagnet was held constant at 80\%. The magnetic field was first rotated by -20 turns, which introduced (-) plectonemes and reduced the extension of the DNA tether. With particle tracking at a frame rate of 40 Hz, the turns of the field were increased in increments of 0.2 turns every 3 seconds up to +20. At each increment, the \textit{z} positions of beads were determined in 120 successive images (blue points). The distance between a tethered and a stuck bead were calculated and plotted (Fig. 9, blue points) along with a moving average (orange).
Re-winding the DNA eliminated (-) plectonemes until the molecule was fully relaxed and extended to a maximum. Further winding beyond the natural twist of the double-helix induced (+) plectonemes which reduced the extension of the tether again. The slight peak near zero turns and the periodic modulation in the blue points indicates that the bead was tethered by two DNA molecules anchored near to each other on the bead and the glass. Overall, the modulation of plectoneme formation verifies proper modulation of the orientation by the electromagnet without interfering with particle tracking such that the slight oscillation due to precession of a doubly tethered molecule and the discrete length decrease that distinguishes the first twist of a double tether were easily observed. At the 100 mM NaCl salt concentration of stretching buffer, the symmetry of the curve in Figure 9 indicates an tension of about 0.5 pN.\cite{strick1996elasticity}
\begin{figure}[t!]
\includegraphics[scale=0.6]{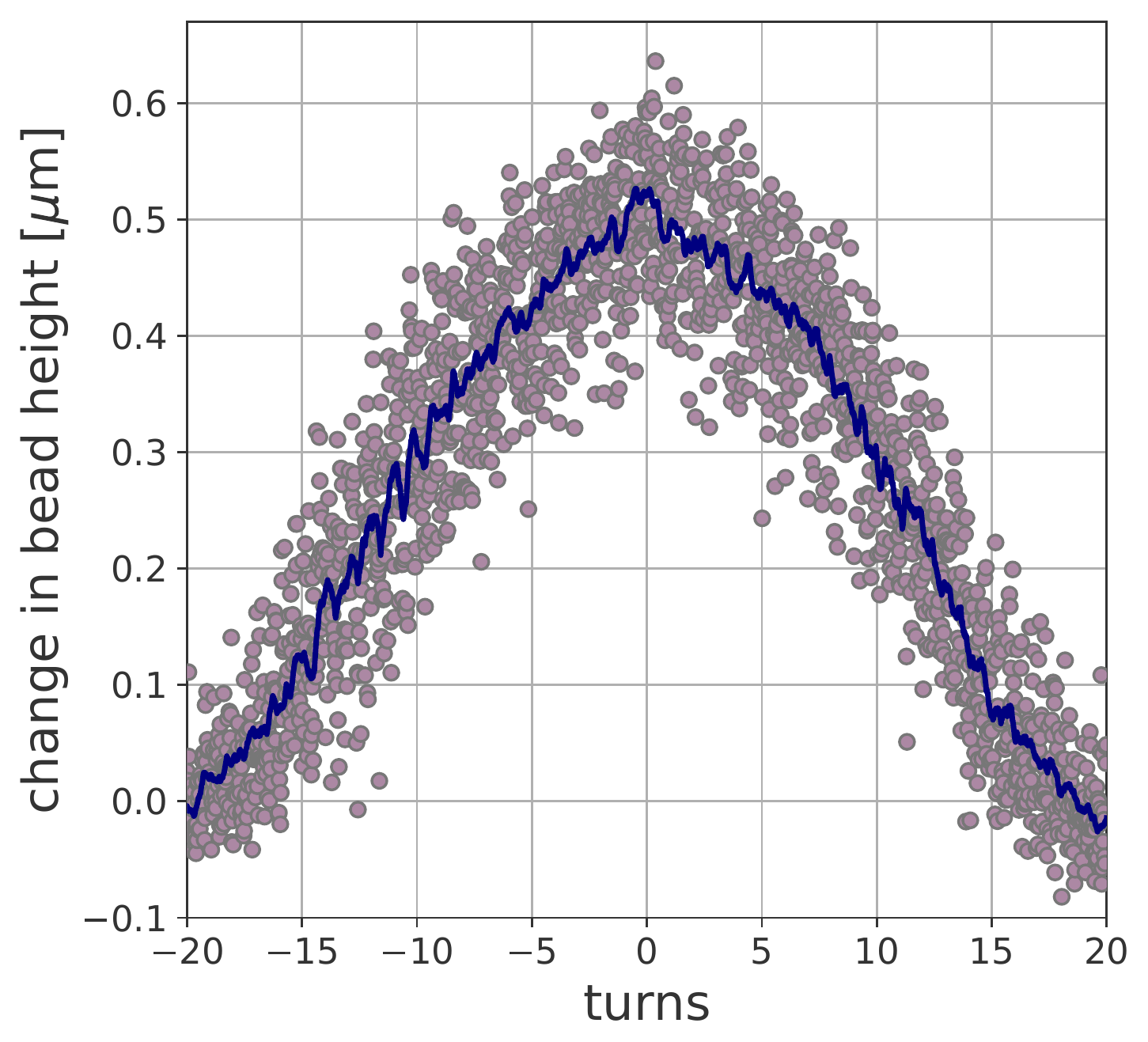} % Here is how to import EPS art
\caption{\label{fig_5} \textbf{A plot of the relative  \textit{z} extension of a DNA tether versus applied twist.} The extension of the DNA tether, \textit{z} position of the bead, (\textit{blue} data and \textit{orange} moving average) decreased as the DNA was wound or unwound. The \textit{z} position at -20 turns was set to zero.} 
\end{figure}
\subsubsection{\label{StepUpDown} Rapid force loading}
Tension along the double helix modulates the twist-writhe partitioning of a DNA molecule and influences energetically favorable conformations. For investigating conformational changes, an instrument capable of abrupt force modulation with high temporal resolution is desirable. In an electromagnet, the magnetic field can be suddenly changed by altering the current with no mechanical motion required. With the pole pieces at a height of 0.35 mm the strength setting was raised from 0\% to 80\%, held at that value for 15 seconds, and lowered again to 0\%. The effective tether length versus time plot is shown in Figure 10. The average forces calculated for the first and second 0\% current intervals are statistically indistinguishable, with wide ranging effective tether lengths as expected at low forces which allow random conformational changes. In contrast, DNA tether lengths were tightly grouped when the force was suddenly increased to 14.7 pN. The transitions between low and high force steady states required less than 1 second with no spurious tether lengths that would indicate particle tracking errors. Note however the non-zero magnetic fields at 0\% current, which indicate remnant fields in the pole pieces.
\begin{figure}[h]
\includegraphics[scale=0.6]{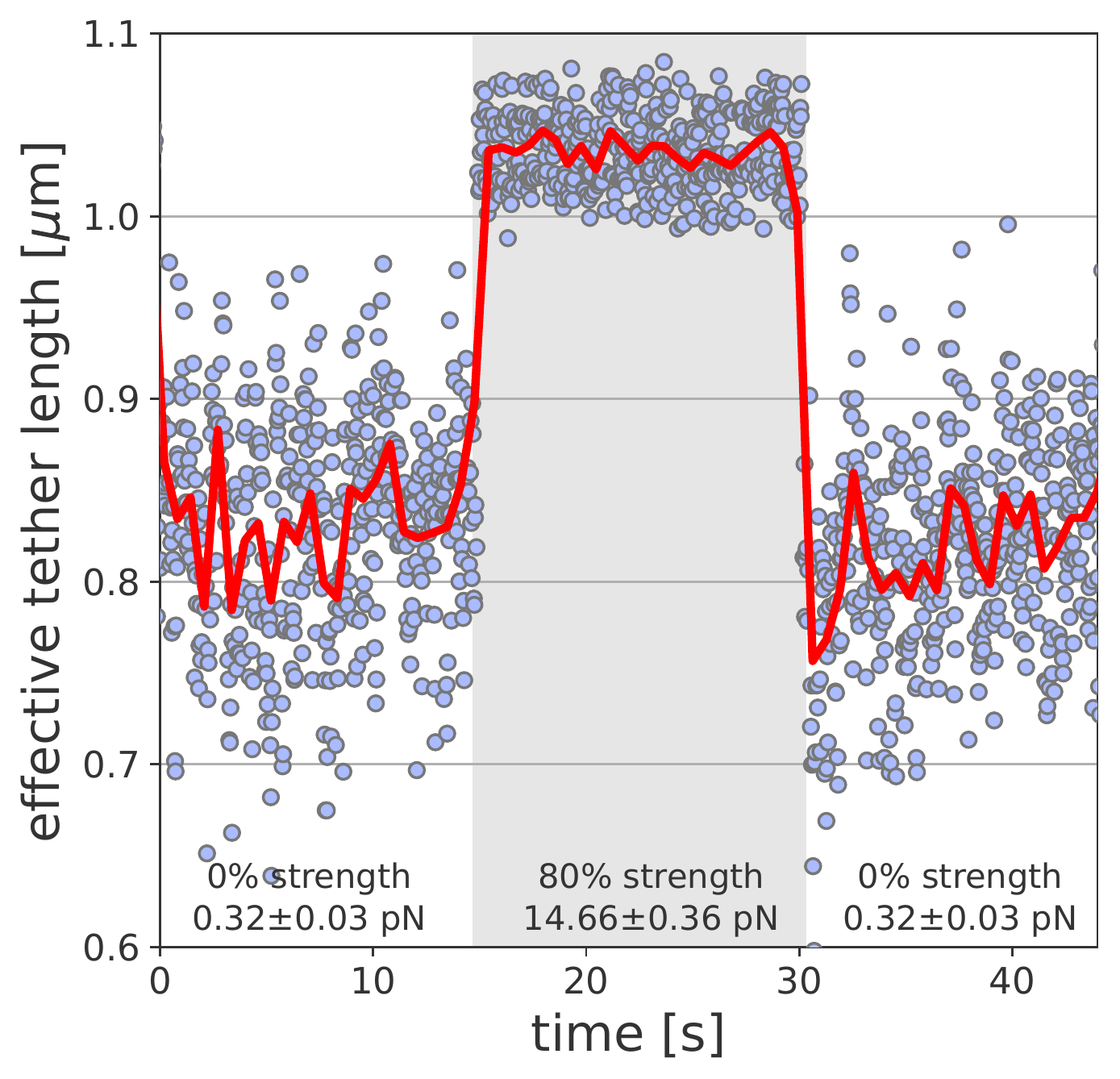} % Here is how to import EPS art
\caption{\label{fig_6} \textbf{A plot of the relative extension of a DNA tether versus time.} The DNA tether length in a magnetic tweezing experiment at a frame rate of 40 Hz oscillated around 0.85 microns under 0.32 pN of force initially. At 15 seconds, the extension of the DNA tether (\textit{blue} data and \textit{red} moving average) increased abruptly to greater than 1 micron as the current was stepped to 80\% and the force  increased to 14.66 pN. Fifteen seconds later, the current was cut back to 0\% and the DNA tether length fell back to the starting level within 12.5 ms. }
\end{figure}
\subsubsection{\label{Min_hyst} Minimal hysteresis}
The remnant fields noted in Fig. 10, emphasize that materials, like the low carbon steel in the solenoid cores, are not perfectly paramagnetic. Once aligned by an externally applied field, magnetic domains in the material may not randomize completely when the field strength is changed, leaving a remnant field. Such hysteresis might prevent the actual magnetic field from returning to previously set values especially after large changes. Therefore, a step-wise test was conducted starting at an initial value (0\%) for a 15 second interval and then raising the current abruptly in steps of 20, 40, 60, 80, or 100\%, followed after another 15 second interval by a return to the initial current. As shown in Fig. 11, abrupt increases in current created commensurately large increases in force on the bead. Comparison with the magnitudes of forces produced by more incremental changes as shown in Fig. 8 reveals no remarkable hysteresis on this time scale. For instance a change from 0 to 100\% generated 18 pN of force in either mode. However, the standard deviation of the force values increased slightly for larger changes indicating non-zero hysteresis that may affect large, rapid force increases. Upon abruptly decreasing the current, the force dropped repeatedly to near baseline values with small standard deviations equal to the initial value regardless of the magnitude of the drop current. This step-wise test indicates no severe limitations associated with abruptly modulating the field strength with the electromagnet.  
\begin{figure}[h]
\includegraphics[scale=0.4]{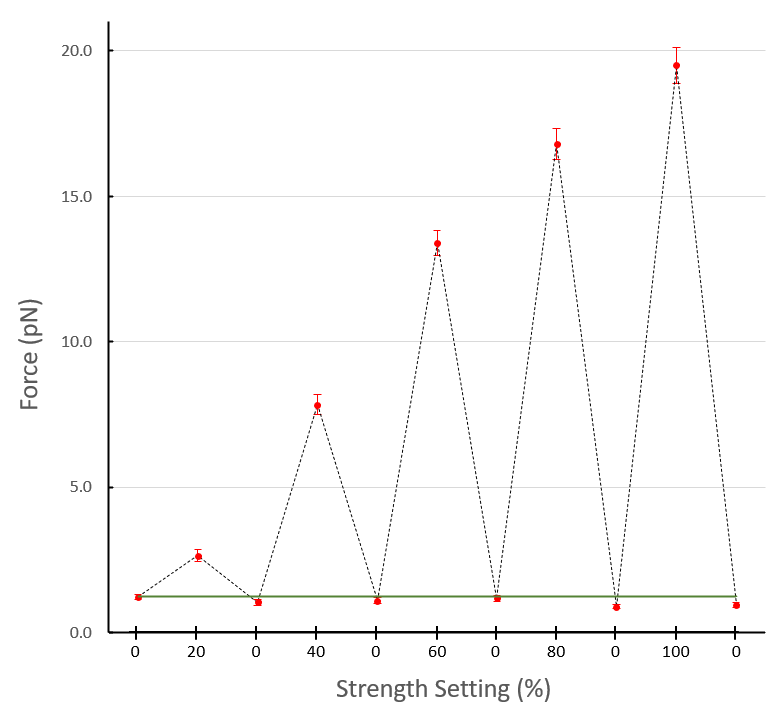} % Here is how to import EPS art
\caption{\label{fig_7} \textbf{A Step-wise Test of Hysteresis:} The calculated force (red) at each current setting is similar to those produced by more incremental increases (see Fig. 8). After abrupt current increases to generate higher forces, the suddenly dropping the current to zero restored the initial force value at 0\% strength (green).} 
\end{figure}
\subsubsection{\label{conclusion} Conclusion}
An electromagnet that exerts up to 20 pN of force on super-paramagnetic beads was realized on the basis of a conceptually simple design. It produces an ample range of physiologically relevant forces on one-micron-diameter, paramagnetic beads, and tenfold changes of the applied force can be achieved in ten ms or less without disturbing particle tracking. These features will facilitate single molecule investigations of polymers in biophysics.
\begin{acknowledgments}
We are grateful to Derrica McCalla for synthesis of the DNA construct.  We also would like to thank the MakeEmory for access to electronic prototyping equipment. This work was supported by a grant from the National Institutes of Health to L.F. (R01 GM084070)
\end{acknowledgments}
\section{References}
\bibliography{aipsamp}% Produces the bibliography via BibTeX.

%\appendix

%\section{Numerical modelling of electromagnetic system}

%Our design

%\section{Electrical hardware}

%\subsection{A subsection in an appendix}

\end{document}